\documentclass[hyperref]{cernrep}
\usepackage{xcolor}
\usepackage[caption=false]{subfig}

\hypersetup{colorlinks=true,linkcolor=blue,citecolor=blue,
urlcolor=blue,	pdfauthor={Aleksas Mazeliauskas}}

\newcommand*{\pp}{$pp$}
\newcommand*{\pO}{$p$O}
\newcommand*{\OO}{OO}
\newcommand*{\pPb}{$p$Pb}
\newcommand*{\PbPb}{PbPb}
\newcommand*{\RAA}{$R_\text{AA}$}
\usepackage[sort&compress,numbers]{natbib}
\begin{document}
\title{Opportunities of OO and $p$O collisions at the LHC}
\author{Jasmine Brewer, Aleksas Mazeliauskas and Wilke van der Schee}
\institute{Theoretical Physics Department, CERN, 1211 Geneva 23, Switzerland}

\begin{abstract}
This is the summary document of the virtual workshop "Opportunities of OO and \pO\ collisions at the LHC", which took place Feb 4-10, 2021. We briefly review the presented  perspectives on the physics opportunities of oxygen--oxygen  and proton--oxygen collisions. The full workshop program and the recordings are available at \href{http://cern.ch/OppOatLHC}{cern.ch/OppOatLHC}.
\end{abstract}

\maketitle

The 5-day workshop consisted of 27 invited talks and 8 discussion sessions covering 
the technological feasibility of oxygen beams at the LHC, theoretical descriptions of the initial state, soft dynamics and hard probes in small colliding systems, experimental opportunities with oxygen-oxygen (OO) and proton-oxygen (\pO) collisions, and the impact of \pO\ data on cosmic ray physics. The aim of the workshop was to review the new developments since the CERN Yellow report~\cite{Citron:2018lsq}, which established the scientific case for a short programme with oxygen beams in LHC Run 3\footnote{Projections for $\sqrt{s_\text{NN}}=200\,\text{GeV}$ \OO\ runs at RHIC were also presented at the workshop, though these runs are subject to machine availability. %
}.

With nearly 400 registered participants and on average 186 unique connections per day,
the workshop attracted a wide community interest. In the following we, the organisers, present a brief summary of selected results that were shown during the workshop and highlight some important open questions that were raised during the discussion sessions. Although we attempt to present the consensus view, this document is not officially endorsed by any of the speakers, participants, or the experimental collaborations. We refer to the recorded presentations, which are available at \href{http://cern.ch/OppOatLHC/timetable}{cern.ch/OppOatLHC/timetable}.

\nocite{AlemanyFernandez:2751158,Bruce:2751162,Gagliardi:2751163,Broniowski:2751165,Schlichting:2751166,Zhao:2751276,Shen:2751277,Niemi:2751279,Nijs:2751280,Kanakubo:2751281,Paakkinen:2751491,Huss:2751492,Apolinario:2751493,Noronha-Hostler:2751494,Xie:2751496,Vislavicius:2751544,Altsybeev:2751545,Sickles:2751546,Murray:2751547,Dembinski:2751548,Graziani:2751549,Menjo:2751672,Tiberio:2751673,Pierog:2751915,Perepelitsa:2751916,Li:2752019,Bierlich:2752020}
\section{Oxygen ions at the LHC accelerator complex}
\label{sec:beams}
Oxygen ions have not been previously injected at the LHC, however they are already produced in the lead ion source and can be selected for acceleration at LINAC3~\cite{Citron:2018lsq, AlemanyFernandez:2751158}.
Because lighter ions have heightened radiological impact, accelerating O$^{4+}$ at LINAC3 and LEIR is the preferred option. An oxygen test will be conducted May 25-29, 2021 to
assess experimentally the radiological
impact in the injector complex and evaluate
the necessary mitigation measures. In total 16 tasks remain to be done in the injector complex before the oxygen run. 
A number of  oxygen beam parameters  are currently unknown  (SPS space charge limit, emittance, bunch intensity, transmutation of  ${}^{16}_8$O into other ions, etc.) and the actual performance could be better (or worse) than quoted below.  The  pilot oxygen run is therefore very useful in assessing the performance in the injectors and the LHC, also in view of high-intensity light-ion runs in Run 5.

Subject to the uncertainties above and good machine availability, a tentative 6-8 day schedule of an OO and \pO{} pilot run was outlined (O$p$ not requested)~\cite{Bruce:2751162}. Without changing the beam rigidity ($\sqrt{s_\text{NN}}(\text{OO})=7\,\text{TeV}$, $\sqrt{s_{\text{NN}}}(p\text{O})=9.9\,\text{TeV}$) it would take 2-3  days of OO commissioning and 1 day of physics run to reach the target luminosity $\mathcal{L}_\text{OO}\approx 0.5\,\text{nb}^{-1}$ for ALICE, ATLAS and CMS. It would take 0.5-1 days of \pO{} commissioning and 2.5-3 days of physics run to reach the target luminosity of   $\mathcal{L}_\text{\pO}\approx 2\,\text{nb}^{-1}$ (ATLAS/LHCf and LHCb) ($\mathcal{L}_\text{\pO}> 5\,\text{nb}^{-1}$ for ALICE and CMS). LHCb minimum physics goals with light hadrons might be met with $\mathcal{L}_\text{\pO}\approx 0.2\,\text{nb}^{-1}$ target luminosity~\cite{Citron:2018lsq,Bruce:2751162}, while the LHCf minimum physics program requires $\mathcal{L}_\text{\pO}\approx 0.7\,\text{nb}^{-1}$ for the two vertical detector  positions at low pileup ($\mu < 0.02$), which also limits the delivered luminosity for ATLAS~\cite{Tiberio:2751673}.
Colliding oxygen at a lower beam energy that matches lead in Run 3 ($\sqrt{s_\text{NN}}(\text{PbPb})=5.52\,\text{TeV}$, $\sqrt{s_\text{NN}}(p\text{Pb})=8.8\,\text{TeV}$) would extend commissioning by 2-3 days, and data-taking by 1-2 days due to the lower luminosity~\cite{Bruce:2751162}. A high intensity oxygen beam (above $3\cdot 10^{11}$ charges/beam) would require additional commissioning (3 days) and ramp up time (3-4 days). 

An accurate determination  of the absolute beam luminosity is needed for many observables (see Sections~\ref{sec:hard} and \ref{sec:cosmic}). Luminosity precision of 1.5-3\% could be achieved by van der Meer (vdM) scans for \OO{} and \pO{}  collisions, which take 2-3 hours per experiment (although other experiments can continue collecting data)~\cite{Gagliardi:2751163}.

Precision studies of energy loss depend on the comparison to a \pp\ reference at the same center-of-mass energy and the cancellation of systematic experimental uncertainties.
At the highest beam energies of OO and \pO\ collisions, the \pp\ reference is either unavailable or not precise enough, preventing cancellation of detector-related uncertainties (e.g., $7\,\text{TeV}$ \pp\ data collected in 2010).
However, \pp\ references are planned to be taken in Run 3 at \PbPb\ and \pPb\ center-of-mass energies~\cite{Citron:2018lsq}. Therefore the limitations of different scenarios (taking a \pp\ reference at OO/\pO{} energies, an \OO/\pO\ run at lower energy, or an interpolated reference from PbPb/$p$Pb references) was a subject of intense debate at the workshop and is discussed further in Section~\ref{sec:hard}.

\begin{figure}
\center
\subfloat[\label{fig:ALICEvn}]{ \includegraphics[width=.3\textwidth]{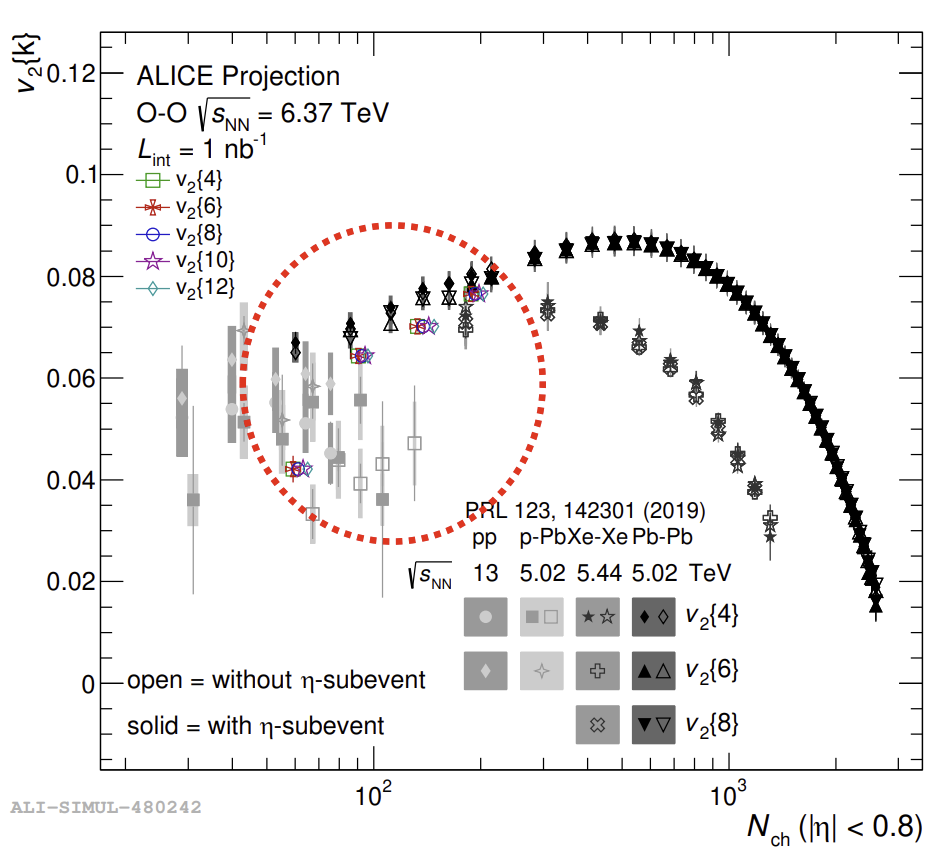}}
\subfloat[\label{fig:v2_meanpT}]{\includegraphics[width=.31\textwidth]{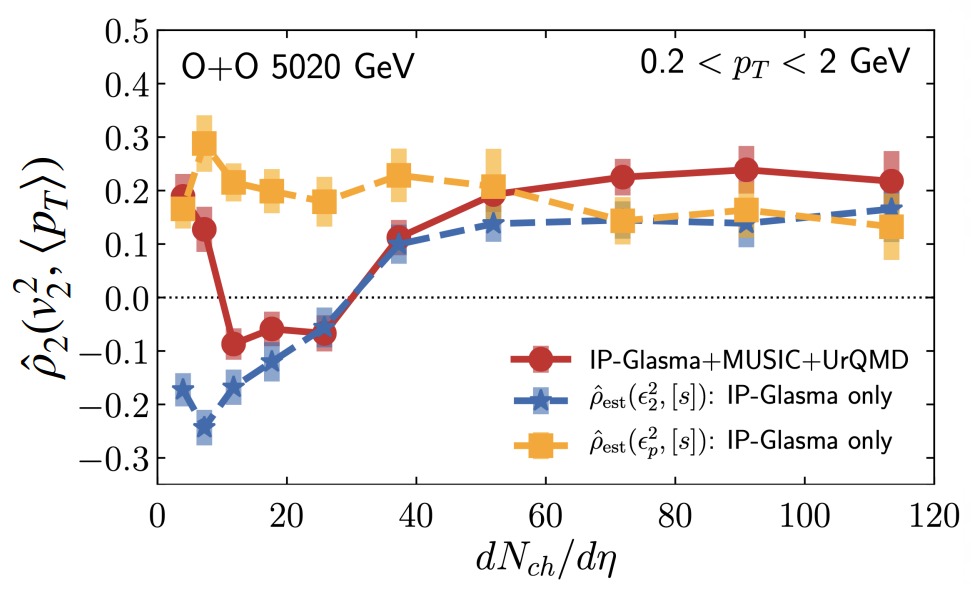}}
\subfloat[\label{fig:alphas}]{ \includegraphics[width=.36\textwidth]{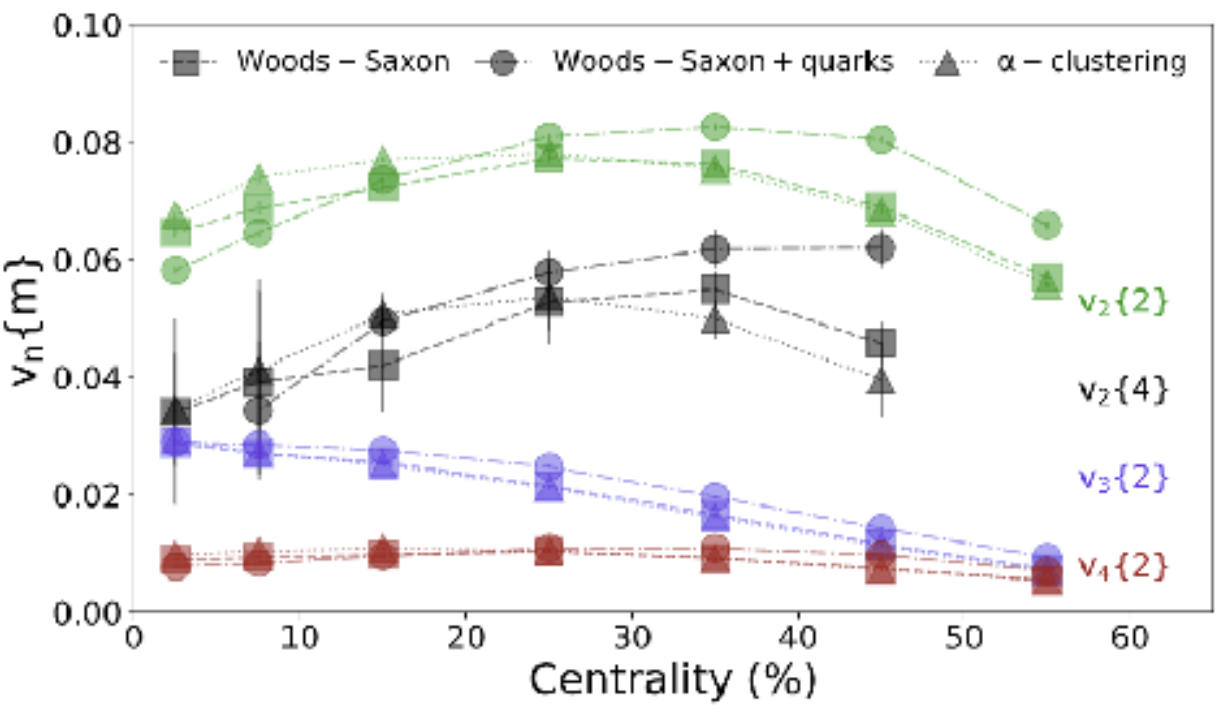}}
\caption{
(a) ALICE experimental projections for anisotropic flow coefficients $v_2\{m\}$ as a function of charged particle multiplicity $N_\text{ch}$ in \OO{}~\cite{Altsybeev:2751545}. 
(b) Correlation between elliptic flow magnitude $v_2^2$ and mean transverse momentum $[p_T]$~\cite{Shen:2751277}. 
(c) Hydrodynamic model predictions for anisotropic flow coefficients $v_n\{m\}$ using a standard Wood-Saxon geometry, a Wood-Saxon model with nucleonic substructure (labelled quarks), and a model where oxygen is a cluster of four alpha particles~\cite{Noronha-Hostler:2751494}. 
}\label{plot:soft}
\end{figure}

\section{Soft QCD dynamics of small systems}%
\label{sec:soft}

Long-range multiparticle correlations, quantified through the Fourier coefficients $v_n$ of transverse momentum anisotropy, have been a hallmark of hydrodynamic behaviour of the quark-gluon plasma (QGP) in heavy-ion collisions. Similar signatures have been observed in peripheral \PbPb, \pPb, and high-multiplicity \pp\ collisions where hydrodynamic behaviour was not expected~\cite{Zhao:2751276}.
The transition of these phenomena from large to small systems can be studied in \OO\ collisions, which are intermediate in multiplicity between \pPb\ and \PbPb{}~\cite{Schlichting:2751166,Altsybeev:2751545}.
ALICE experimental projections demonstrate access to a wide variety of high precision flow measurements in a short run of \OO\ with integrated luminosity $\mathcal{L}_\text{OO}\sim 1\,\text{nb}^{-1}$ \cite{Altsybeev:2751545}.

\OO\ and \pO\ collisions have distinctive features compared to \pPb\ and \PbPb\ that put unique constraints on hydrodynamic models~\cite{Shen:2751277, Nijs:2751280, Niemi:2751279}.
For example, the inclusion of \OO\ collision data in Bayesian analyses appears to improve constraints on QGP parameters compared to only \PbPb\ data \cite{Nijs:2751280}. 
\OO\ collisions have less eccentric geometry than \PbPb\ but a similar number of participating nucleons at the same multiplicity \cite{Shen:2751277, Niemi:2751279}. Since $v_2$ is thought to be driven by geometry and $v_3$ by fluctuations, this suggests that $v_2(\text{OO}) < v_2(\text{PbPb})$, but $v_3(\text{OO}) \sim v_3(\text{PbPb})$.
\OO\ collisions are also more compact at the same final multiplicity compared to \PbPb\ and are expected to have higher medium temperature, which can enhance thermal photon radiation by a factor of $2$ \cite{Noronha-Hostler:2751494, Shen:2751277}.
Compared to \pPb\ the geometry of \OO\ collisions is under better control, since flow coefficients are less sensitive to poorly-constrained spatial fluctuations in the proton, and centrality is more correlated with initial eccentricity \cite{Shen:2751277}.  
ALICE projects to measure a variety of multiparticle correlations in Run 3 in \OO, including $v_n$ measurements for up to $12$ particles \cite{Altsybeev:2751545} (see Fig.~\ref{fig:ALICEvn}), which have substantially decreased sensitivity to jet-like correlations. %
Complementary measurements of $v_2\{4\}$ in \OO\ at lower center-of-mass energy $\sqrt{s_\text{NN}}=200\, \text{GeV}$  may be done by the STAR experiment at RHIC in 2021~\cite{Li:2752019}.

It is a hotly debated question whether the origin of $v_n$ is the same in small and large systems~\cite{Zhao:2751276}.
Initial state correlations and non-equilibrium effects may play an increasingly-important role in small systems ($d N_\text{ch}/d\eta \lesssim 10$) due to their short lifetime \cite{Shen:2751277,Schlichting:2751166}. Model predictions for \OO\ and \pO\ were shown at the workshop based on a  hydrodynamic medium \cite{Shen:2751277, Niemi:2751279} and based on event generators without a deconfined medium \cite{Bierlich:2752020, Schlichting:2751166}.
In small systems, $v_n$ can be sensitive to hadronization and to residual jet-like correlations \cite{Schlichting:2751166}.
A sign change in the multiplicity-dependence of the $v_2^2$-mean $p_T$ correlator at $d N_\text{ch}/d\eta \sim 10$ was proposed as an observable handle on the transition between regimes where initial state correlations or hydrodynamics are dominant \cite{Shen:2751277} (see Fig.~\ref{fig:v2_meanpT}). ALICE projections show this measurement would be possible with $\mathcal{L}_\text{OO}\sim 1\,\text{nb}^{-1}$, though the influence of jet-like correlations requires further study~\cite{Altsybeev:2751545, Li:2752019}. %

Nucleons in $^{16}$O are expected to be clustered into four alpha particles and it is unknown whether this structure may be observable in oxygen collisions. 
Model calculations suggest that eccentricities in \OO\ depend weakly on alpha clustering of nucleons, yielding only few-percent effects in the most central events compared to smooth nuclear distributions \cite{Broniowski:2751165}. Hydrodynamic simulations suggest that sub-nucleonic structure plays a more important role in generating flow harmonics~\cite{Shen:2751277,Noronha-Hostler:2751494} (see Fig.\ref{fig:alphas}).
Eccentricities may depend more strongly on alpha clustering in collisions of oxygen with a heavy nucleus \cite{Broniowski:2751165}  (e.g. with Pb beam and O as a fixed target at LHCb/SMOG2~\cite{Graziani:2751549}).

\begin{figure}
\center
\subfloat[\label{fig:strangeness}]{ \includegraphics[width=.28\textwidth]{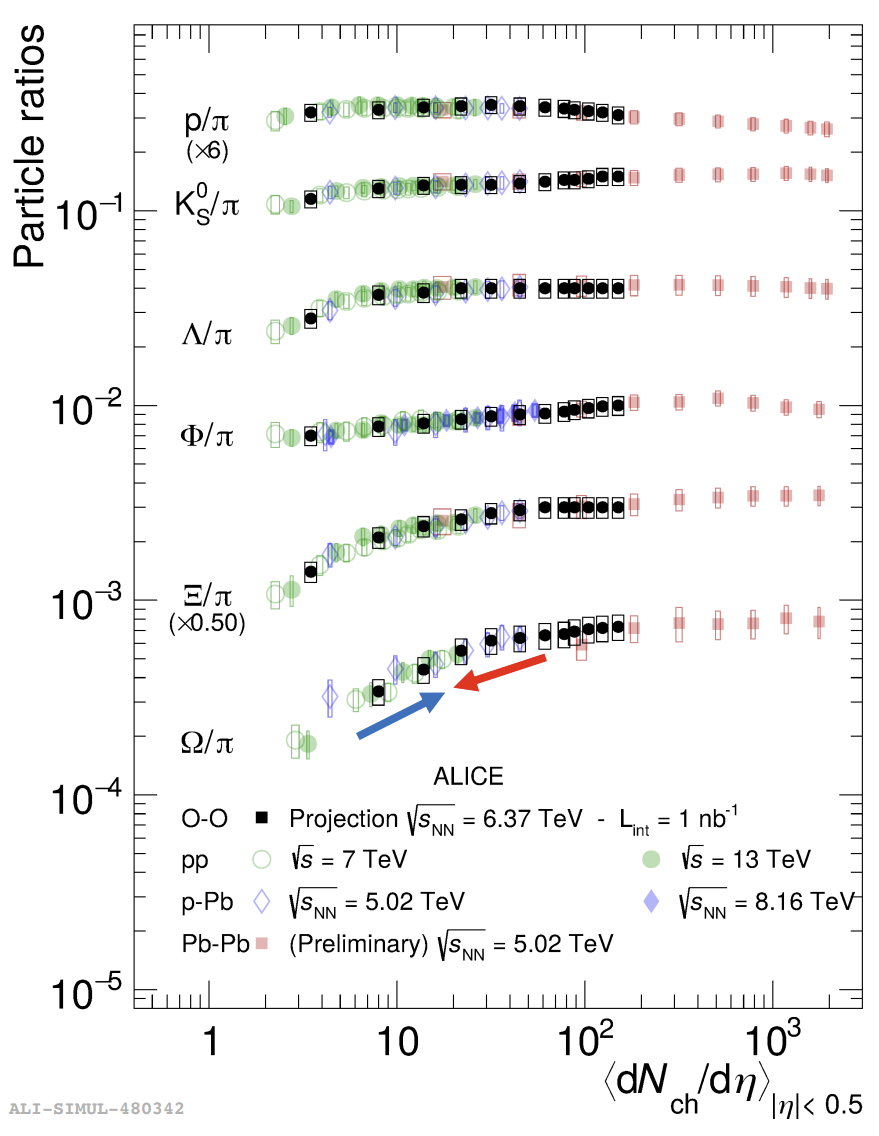}}
\subfloat[\label{fig:hadronRAA}]{
\includegraphics[width=.33\textwidth]{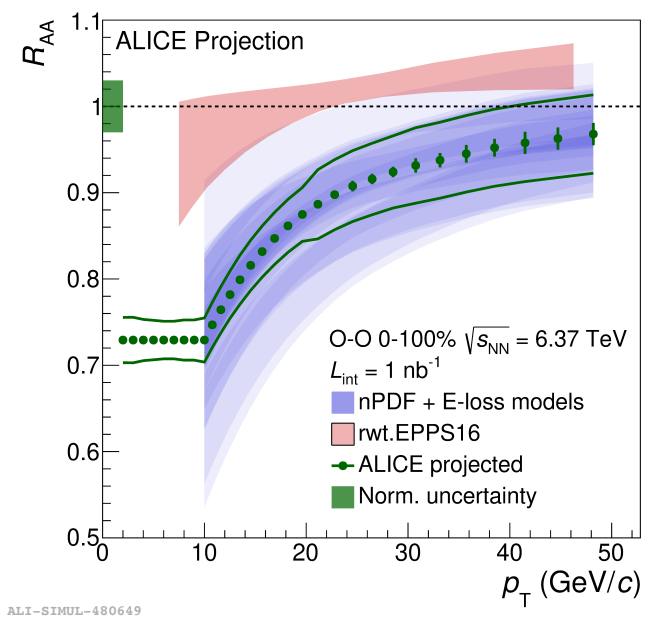}}
\subfloat[\label{fig:RFB}]{ \includegraphics[width=.29\textwidth]{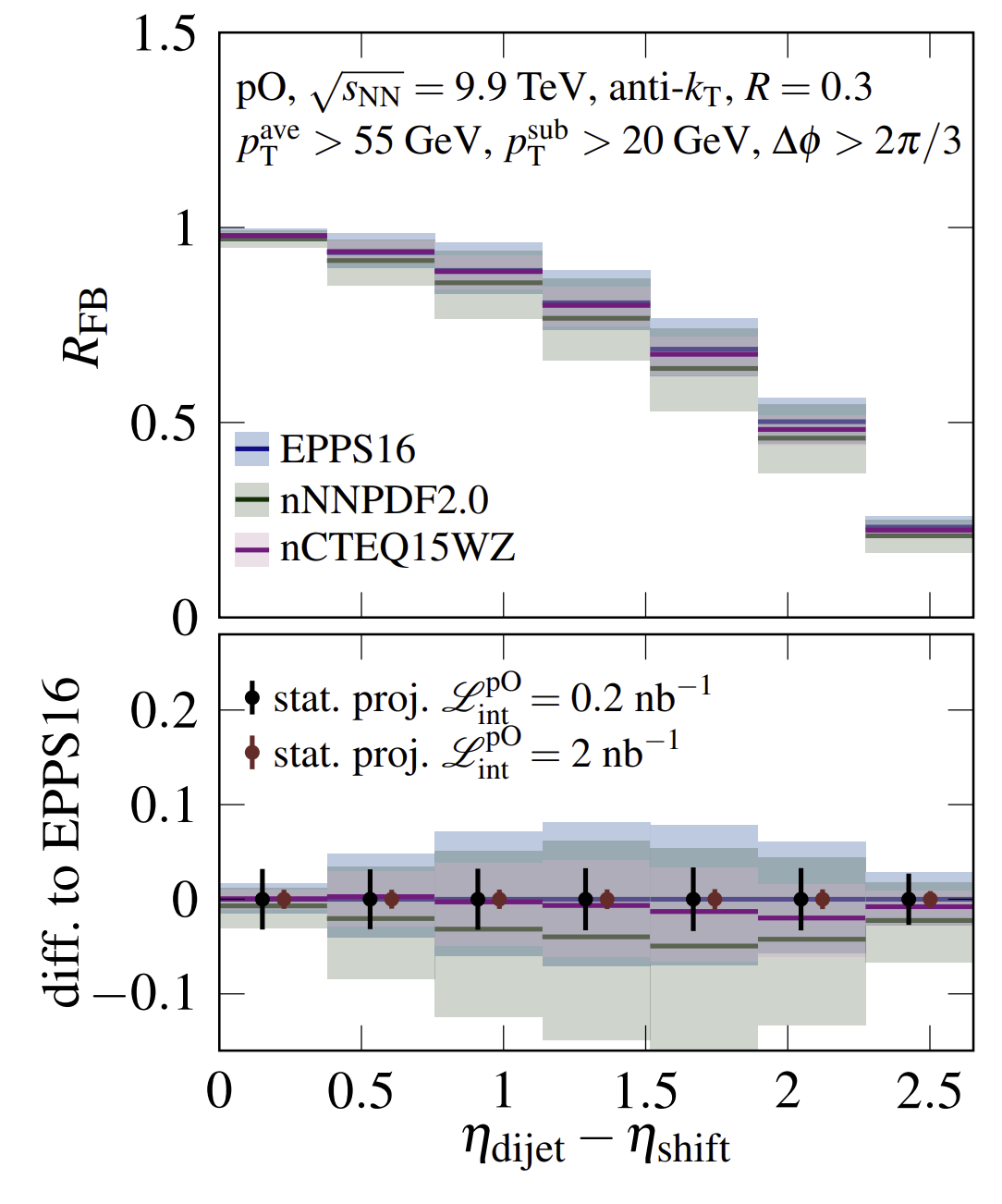}} 
\caption{(a) ALICE projections for strange hadron yields in \OO\ compared to $pp$, \pPb\ and \PbPb\ \cite{Altsybeev:2751545}.  (b) ALICE projections of the charged hadron nuclear modification factor \RAA\ for minimum bias \OO\ compared to energy loss models and the no-energy-loss baseline (nPDFs including uncertainties)~\cite{Vislavicius:2751544}. (c) The forward-backward ratio $R_{\rm FB}$ of dijets, shown for three different nPDF sets, compared to statistical uncertainties with $\mathcal{L}_{p\text{O}}=0.2\,\text{nb}^{-1}$ and $\mathcal{L}_{p\text{O}}=2\,\text{nb}^{-1}$ integrated luminosity~\cite{Paakkinen:2751491}.}
\label{plot:hard}
\end{figure}

\OO\ collisions give a unique opportunity to cover a multiplicity range overlapping with both \pPb{} and \PbPb{} in a single system.
ALICE will be able to measure the production of strange hadrons (see Fig.~\ref{fig:strangeness}), deuterons, ${}^3$He and hypertritons in \OO{}~\cite{Altsybeev:2751545}. %
Finally, ALICE showed that femtoscopy measurements in \OO\ can provide novel insight on the strong interaction potential between hadrons. \OO\ collisions have a larger hadron source size of $2\text{-}3\,\text{fm}$ compared to $1\,\text{fm}$ for \pp{}, which makes it possible to study coupled-channel contributions for $p\text{--}K$ pairs, and the possible existence of a $p\text{--}\Omega^-$ bound state~\cite{Altsybeev:2751545}.

\section{Hard probes in \OO{}  and \pO{}}
\label{sec:hard}

If a dense QCD medium is produced in hadronic collisions, energetic partons can rescatter with it and lose energy. Therefore high-$p_T$ particle modification is an expected
consequence of the final state rescatterings that manifest themselves as hydrodynamic collectivity in the soft sector (see Section~\ref{sec:soft}).
It is an open question why hard probes measurements in small systems, such as \pPb{}, are consistent with no energy loss \cite{Sickles:2751546, Apolinario:2751493}. ALICE, ATLAS, CMS, and LHCb collaborations expect to take a variety of measurements in \OO{} and \pO{} to address high-$p_T$ rescattering in small systems~\cite{Vislavicius:2751544,Sickles:2751546,Murray:2751547,Dembinski:2751548}.
The sPHENIX experiment at RHIC could also measure hard probes at $\sqrt{s_\text{NN}}=200\text{ GeV}$ if a window of opportunity arises for an \OO\ run after 2025~\cite{Perepelitsa:2751916}.

One way to quantify the modification of high-$p_T$ particles and jets is through the nuclear modification factor \RAA, which is the ratio of the cross section for high-$p_T$ processes in a nuclear environment to a (scaled) \pp\ cross section. 
\RAA{} differs from unity due to the nuclear modification of parton distribution functions (nPDFs) and medium induced energy loss~\cite{Apolinario:2751493}.
Model calculations show that energy loss signal may be small in small systems ($5\text{--}10\%$ for hadrons with $p_T > 20 \text{ GeV}$ in \OO~\cite{Huss:2751492}), therefore precise theoretical and experimental control of observables is crucial. 
Minimum bias \OO\ collisions do not have centrality selection biases like peripheral \PbPb{}, though vdM scans are necessary to determine the \RAA{} normalization~\cite{Vislavicius:2751544}. 
ALICE estimates a systematic uncertainty of about $5\%$ (statistical uncertainties subdominant) on the minimum bias hadron \RAA\ measurement, which could distinguish energy loss models from the nPDF baseline (see Fig.~\ref{fig:hadronRAA}). Similar uncertainties were estimated for centrality-selected \RAA{} but with possible selection bias. ATLAS showed projections for $<5\%$ statistical uncertainties on jet ($50\,\text{GeV}/c<p^j_T<240\,\text{GeV}/c$) and hadron ($p^h_T<80\,\text{GeV}/c$) \RAA{} for $\mathcal{L}_\text{OO}\sim 0.5 \text{ nb}^{-1}$~\cite{Sickles:2751546}.

Accurate energy loss measurements could also be possible with coincidence measurements of hadron-jet or photon-jet spectra normalized to the trigger (hadron or photon) multiplicity~\cite{Vislavicius:2751544,Perepelitsa:2751916}. Such self-normalizing measurements can cancel experimental systematic uncertainties and biases.
ALICE projected sensitivity to $\Delta p_T \sim 160\,\text{MeV}$ shifts in hadron-triggered jet spectra in minimum bias collisions, with dominant uncertainty from the \pp\ reference interpolation~\cite{Vislavicius:2751544}. Even smaller shifts might be measured without the need of $pp$ reference by comparing spectra from collisions with high and low event activities,
but precise baseline calculations could be more challenging. 
New results for $Z$-jet and $\gamma$-hadron spectra shown at the workshop indicate that triggered observables can still have non-negligible uncertainties from nPDFs and impact the interpretation of the signal~\cite{Xie:2751496, Huss:2751492}.

The azimuthal momentum anisotropy of high-$p_T$ particles (as measured with respect to low-$p_T$ particles) is typically associated with differential energy loss, but does not require a \pp\ reference.
  ATLAS and ALICE showed projections of high-$p_T$ $v_2$  measurements in \OO~\cite{Sickles:2751546,Altsybeev:2751545}.
ALICE also projected 10\% accuracy for D meson \RAA\  and about 30\% accuracy (statistics dominated) for $v_2$ measurements.
D meson \RAA\ and $v_2$ are sensitive to the energy loss of charm quarks at high $p_T$ and their momentum diffusion at low $p_T$~\cite{Noronha-Hostler:2751494}.

Currently there is no experimental oxygen data available for global nPDF fits and the results for ${}^{16}$O depend on the assumed  parametrization of the $A$-dependence in nPDF fits~\cite{Paakkinen:2751491}. Furthermore, hadronic data (e.g. dijets) is included only for Pb ions at the LHC. Therefore \pO\ data is important for the reliability of nPDFs for light ions. Forward-backward ratios of dijets with  $\mathcal{L}_\text{\pO}\sim 2\, \text{nb}^{-1}$  would already give a valuable anchor on gluon modification in light nuclei  (see Fig.~\ref{fig:RFB}). The constraints could be further strengthened with a reliable \pp\ reference, which would enable the study of more varied $x$-dependence of nPDFs. Since $^{16}$O is isoscalar, 
$Z$ and $W^{\pm}$ measurements in \pO\ could put unique constraints on the strange quark nPDF, but would require considerably more statistics, $\mathcal{L}_\text{\pO}\sim 2 \,\text{pb}^{-1}$.

Finally, accurate measurements of \RAA-type observables rely on having a \pp\ reference to cancel some systematic uncertainties in the spectra.
As discussed in Section~\ref{sec:beams}, obtaining a \pp\ reference at the same center-of-mass energy would require additional run time. On the other hand, interpolating the reference between energies and cancelling sizeable systematic uncertainties can be challenging, especially at high $p_T$~\cite{Sickles:2751546}.
In the 2013 \pPb\ run, the interpolated reference used by ATLAS resulted in large ($\sim$20\%) systematic uncertainties of charged hadron $R_\text{pPb}$ measurement for $p_T > 20 \text{ GeV/c}$.
However, ALICE studied the uncertainties and their cancellation in hadron \RAA\  with an interpolated \OO{} reference. They projected
$\sim 5\%$ uncertainties on hadron \RAA\ for $p_T<50\,\text{GeV}/c$ in \OO\ with $\lesssim 3\%$ uncertainty due to the interpolated reference \cite{Vislavicius:2751544}.

\section{Forward physics and cosmic rays}
\label{sec:cosmic}

Forward kinematics of $\sqrt{s_\text{NN}}=9.9\,\text{TeV}$ \pO\ collisions  corresponds to cosmic rays with energy $>10^{16}\,\text{eV}$ hitting the Earth's atmosphere (mostly  ${}^{14}$N and ${}^{16}$O). Therefore oxygen collisions during the LHC Run 3 will provide valuable  hadronic data for cosmic ray modelling and will help address the outstanding problem that there are more muons in air showers than predicted ~\cite{Menjo:2751672,Tiberio:2751673,Pierog:2751915,Dembinski:2751548}.

\begin{figure}
    \centering
    \subfloat[\label{fig:fig3b}]{ \includegraphics[width=0.32\linewidth]{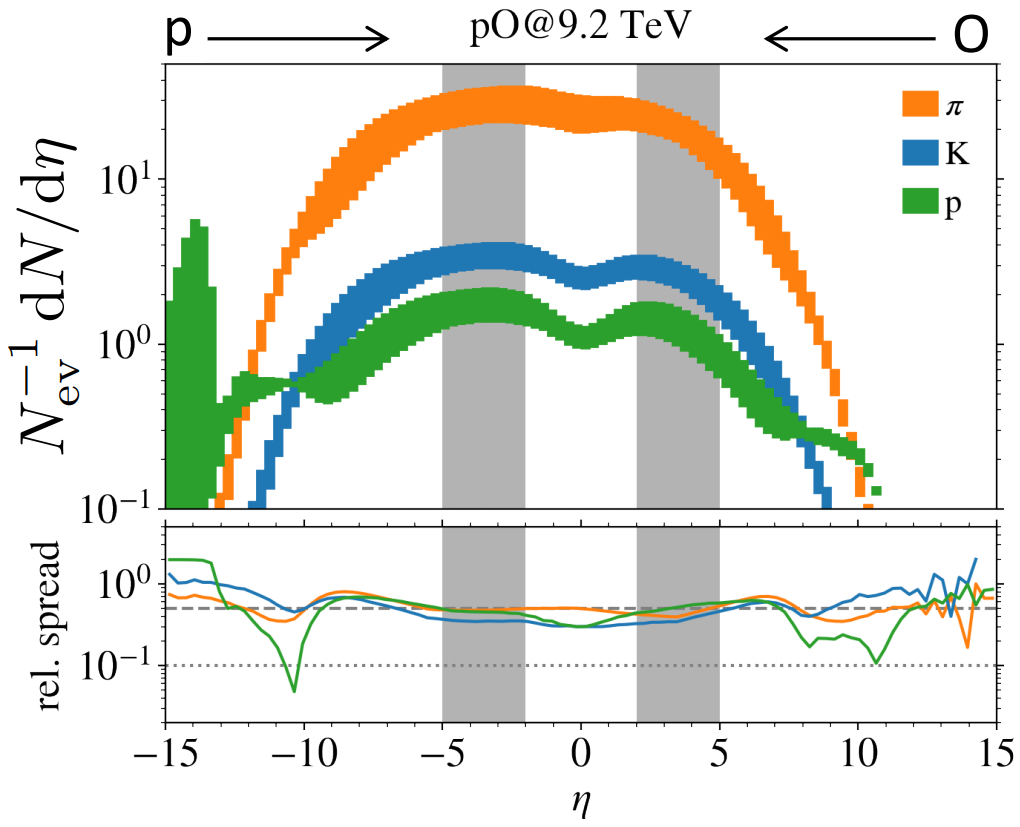}}
    \subfloat[\label{fig:fig3a}]{ \includegraphics[width=0.32\linewidth]{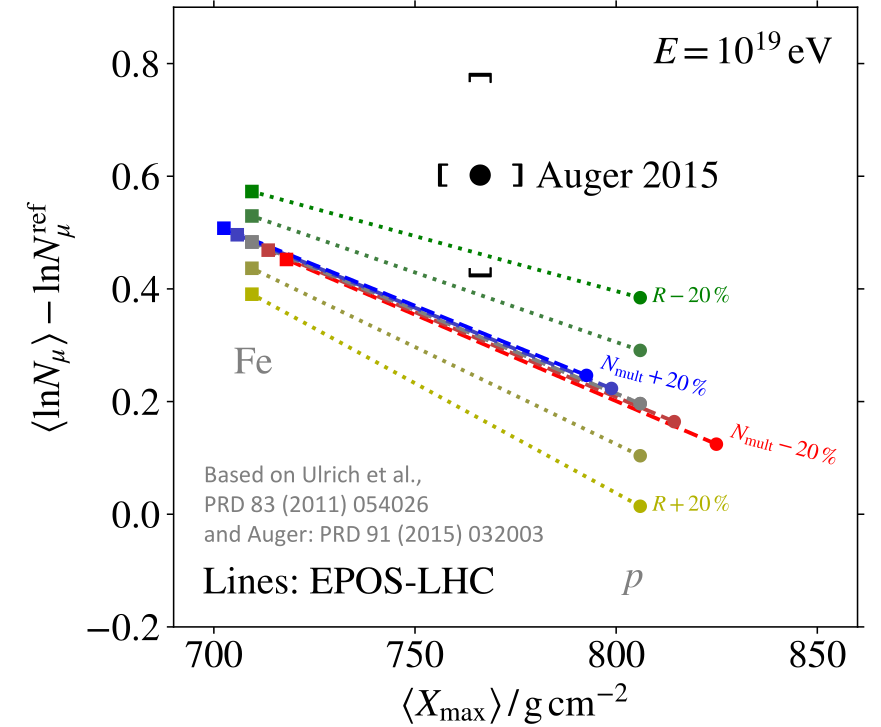}}
    \subfloat[\label{fig:fig3c}]{ \includegraphics[width=0.32\linewidth]{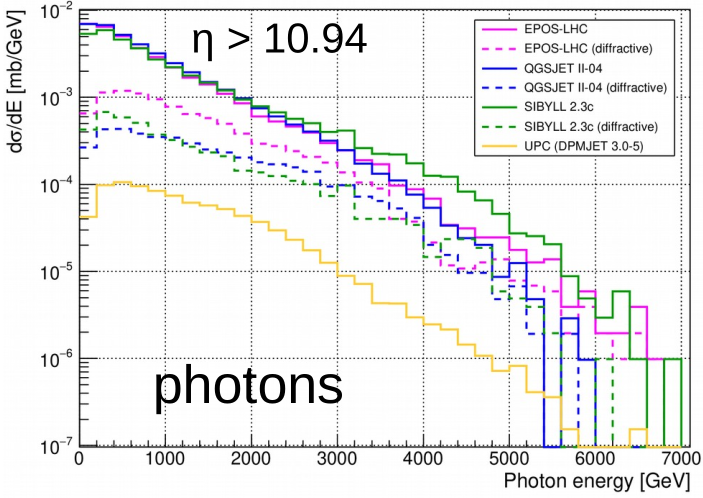}}
    \caption{(a) The spread of identified particle spectra in \pO{} collisions for different models. The gray area indicates LHCb acceptance~\cite{Dembinski:2751548}. (b) Logarithmic muon content as a function of the average shower depth $\left<X_\text{max}\right>$ measured by  Auger observatory and compared to model calculations with different values of the $\pi^0$ energy fraction $R$~\cite{Dembinski:2751548}. (c) Forward photon spectra in diffractive and inelastic collisions for different models compared to the UPC background~\cite{Tiberio:2751673}.}\label{fig:cosmic}
\end{figure}

LHCb is expected to perform at full efficiency for both \OO{} and \pO{} collisions and 
measure identified light-hadron spectra in the acceptance $2<\eta<5$, which can be used to constrain the current 50\% spread among models (see Fig.~\ref{fig:fig3b})~\cite{Dembinski:2751548}.
The precise value of the $\pi^0$ energy fraction $R=E_{\pi^0}/E_\text{other hadrons}$ is particularly important in cosmic ray showers and current observations point to smaller values of $R$ than used in models or measured in $pp$ (see Fig.~\ref{fig:fig3a}). If confirmed, a smaller $R$ value at forward \pO{} kinematics could potentially be attributed to strangeness enhancement--one of the signals of QGP formation. $D$ meson measurements could also help constrain the atmospheric lepton background for astrophysical neutrinos searches~\cite{Dembinski:2751548,Pierog:2751915}.

LHCb plans to simultaneously take fixed-target data at a higher rate with SMOG2~\cite{Graziani:2751549}.
Oxygen collisions on hydrogen taken during the \OO{} run would provide complementary forward rapidity data ($3.9<\eta'<6.9$ in the oxygen rest-frame) at $\sqrt{s_\text{NN}}(p\text{O})=81\,\text{GeV}$. Other configurations of oxygen collisions on different targets would also be possible,  time permitting.

The LHCf forward detectors near the ATLAS interaction point measure
photons, neutral pions and neutrons at very forward rapidities $|\eta|>8.4$, which are most relevant for air shower development~\cite{Menjo:2751672,Tiberio:2751673,Pierog:2751915}. 
Current \pPb\ measurements suffer from large ultra-peripheral collision (UPC) backgrounds. For \pO{} the UPC background is significantly reduced and a combined ATLAS-LHCf analysis would allow to measure diffractive collisions (see Fig.~\ref{fig:fig3c}). Additionally, current neutron data in $pp$ show significant contributions from pion exchange, which is not described by current models.  $\pi$O interactions are particularly important for secondary cosmic ray showers and might be accessible in \pO{} collisions~\cite{Pierog:2751915}.

Finally, LHCf will not be able to operate after Long Shutdown 3 due to infrastructure modification at its current position, and the  Pierre Auger Observatory will finish operations around $2030$. Therefore the cosmic ray community strongly supports
\pO\ data taking at the highest available energies during LHC Run 3~\cite{Menjo:2751672,Tiberio:2751673,Pierog:2751915}.

\section{Conclusions}

Detailed machine, theoretical and experimental presentations  during the virtual workshop "Opportunities of OO and \pO\ collisions at the LHC" have reaffirmed the readiness and wide community interest for oxygen collisions at the LHC. OO and \pO{} collisions during Run 3 will provide unique discovery opportunities and address long standing puzzles across different fields. 
A number of open questions have been identified, 
which will have to be scrutinised further to optimise the physics output from this short run. Undoubtedly, a pilot oxygen run will open a new chapter for the LHC as a light-ion machine.

\section*{Acknowledgements}
We wish to thank speakers, discussion leaders, and all participants for their contributions. We are especially grateful to Michelangelo Mangano and Urs Wiedemann for their help.

\setlength{\bibsep}{3pt}
\bibliographystyle{hunsrtnat}
\bibliography{summary}

\begin{thebibliography}{28}
\expandafter\ifx\csname natexlab\endcsname\relax\def\natexlab#1{#1}\fi
\expandafter\ifx\csname url\endcsname\relax
  \def\url#1{{\tt #1}}\fi

\bibitem[Citron et~al.(2019)]{Citron:2018lsq}
Z.~Citron et~al.
\newblock {Report from Working Group 5}: {Future physics opportunities for
  high-density QCD at the LHC with heavy-ion and proton beams}.
\newblock {\em CERN Yellow Rep. Monogr.}, 7:\penalty0 1159--1410, 2019,
  1812.06772.

\bibitem[Alemany~Fernandez(2021)]{AlemanyFernandez:2751158}
Reyes Alemany~Fernandez.
\newblock {Preparing the CERN Ion Injector Chain for an LHC Oxygen Run}.
\newblock Feb 2021.
\newblock URL \url{https://cds.cern.ch/record/2751158}.

\bibitem[Bruce(2021)]{Bruce:2751162}
Roderik Bruce.
\newblock {LHC machine scenario for a short oxygen run}.
\newblock Feb 2021.
\newblock URL \url{https://cds.cern.ch/record/2751162}.

\bibitem[Gagliardi(2021)]{Gagliardi:2751163}
Martino Gagliardi.
\newblock {Luminosity determination with heavy-ion beams at the LHC}.
\newblock Feb 2021.
\newblock URL \url{https://cds.cern.ch/record/2751163}.

\bibitem[Broniowski(2021)]{Broniowski:2751165}
Wojciech Broniowski.
\newblock {Eccentricities in collisions with 16O and 12C}.
\newblock Feb 2021.
\newblock URL \url{https://cds.cern.ch/record/2751165}.

\bibitem[Schlichting(2021)]{Schlichting:2751166}
Soeren Schlichting.
\newblock {Initial state and non-equilibrium dynamics in small and large
  systems}.
\newblock Feb 2021.
\newblock URL \url{https://cds.cern.ch/record/2751166}.

\bibitem[Zhao(2021)]{Zhao:2751276}
Wenbin Zhao.
\newblock {Collectivity and QGP signals in Large and Small systems}.
\newblock Feb 2021.
\newblock URL \url{https://cds.cern.ch/record/2751276}.

\bibitem[Shen(2021)]{Shen:2751277}
Chun Shen.
\newblock {Dynamical Modeling of the Collectivity in pO and OO Collisions}.
\newblock Feb 2021.
\newblock URL \url{https://cds.cern.ch/record/2751277}.

\bibitem[Niemi(2021)]{Niemi:2751279}
Harri Niemi.
\newblock {Multiplicity and flow in O+O collisions from the EKRT model}.
\newblock Feb 2021.
\newblock URL \url{https://cds.cern.ch/record/2751279}.

\bibitem[Nijs(2021)]{Nijs:2751280}
Govert Nijs.
\newblock {Bayesian Analysis of Oxygen-Oxygen Collisions}.
\newblock Feb 2021.
\newblock URL \url{https://cds.cern.ch/record/2751280}.

\bibitem[Kanakubo and Hirano(2021)]{Kanakubo:2751281}
Yuuka Kanakubo and Tetsufumi Hirano.
\newblock {Hadron yield ratios in dynamical core-corona initialization from
  small to large systems}.
\newblock Feb 2021.
\newblock URL \url{https://cds.cern.ch/record/2751281}.

\bibitem[Paakkinen(2021)]{Paakkinen:2751491}
Petja Paakkinen.
\newblock {Current status of nPDFs and prospects for pO and OO collisions}.
\newblock Feb 2021.
\newblock URL \url{https://cds.cern.ch/record/2751491}.

\bibitem[Huss(2021)]{Huss:2751492}
Alexander~Yohei Huss.
\newblock {Discovering partonic rescattering in light nucleus collisions}.
\newblock Feb 2021.
\newblock URL \url{https://cds.cern.ch/record/2751492}.

\bibitem[Apolinario(2021)]{Apolinario:2751493}
Liliana Apolinario.
\newblock {Jet Quenching from light to dense systems}.
\newblock Feb 2021.
\newblock URL \url{https://cds.cern.ch/record/2751493}.

\bibitem[Noronha-Hostler(2021)]{Noronha-Hostler:2751494}
Jacquelyn Noronha-Hostler.
\newblock {The elusive energy loss signal of the Quark Gluon Plasma in OO
  collisions}.
\newblock Feb 2021.
\newblock URL \url{https://cds.cern.ch/record/2751494}.

\bibitem[Xie(2021)]{Xie:2751496}
Man Xie.
\newblock {$\gamma$-hadron spectra in $p$ + Pb collisions at $\sqrt{s_{\rm
  NN}}=5.02$ TeV}.
\newblock Feb 2021.
\newblock URL \url{https://cds.cern.ch/record/2751496}.

\bibitem[Vislavicius(2021)]{Vislavicius:2751544}
Vytautas Vislavicius.
\newblock {ALICE goals and projections for hard probes measurements}.
\newblock Feb 2021.
\newblock URL \url{https://cds.cern.ch/record/2751544}.

\bibitem[Altsybeev(2021)]{Altsybeev:2751545}
Igor Altsybeev.
\newblock {ALICE goals and projections for flow and hadron
  production/interaction measurements}.
\newblock Feb 2021.
\newblock URL \url{https://cds.cern.ch/record/2751545}.

\bibitem[Sickles(2021)]{Sickles:2751546}
Anne~Marie Sickles.
\newblock {ATLAS contribution}.
\newblock Feb 2021.
\newblock URL \url{https://cds.cern.ch/record/2751546}.

\bibitem[Murray(2021)]{Murray:2751547}
Michael Murray.
\newblock {Physics opportunities with oxygen collisions in CMS}.
\newblock Feb 2021.
\newblock URL \url{https://cds.cern.ch/record/2751547}.

\bibitem[Dembinski(2021)]{Dembinski:2751548}
Hans~Peter Dembinski.
\newblock {Oxygen beams and LHCb: prospects of pO and OO collisions for nuclear
  and astroparticle physics}.
\newblock Feb 2021.
\newblock URL \url{https://cds.cern.ch/record/2751548}.

\bibitem[Graziani(2021)]{Graziani:2751549}
Giacomo Graziani.
\newblock {Oxygen beams and LHCb: prospects of collisions with fixed-targets}.
\newblock Feb 2021.
\newblock URL \url{https://cds.cern.ch/record/2751549}.

\bibitem[Menjo(2021)]{Menjo:2751672}
Hiroaki Menjo.
\newblock {LHCf achievements at pp and pPb}.
\newblock Feb 2021.
\newblock URL \url{https://cds.cern.ch/record/2751672}.

\bibitem[Tiberio(2021)]{Tiberio:2751673}
Alessio Tiberio.
\newblock {LHCf motivations and prospects of p-O collisions}.
\newblock Feb 2021.
\newblock URL \url{https://cds.cern.ch/record/2751673}.

\bibitem[Pierog(2021)]{Pierog:2751915}
Tanguy Pierog.
\newblock {Hadronic interactions and air showers : the need of Oxygen beam with
  LHCf}.
\newblock Feb 2021.
\newblock URL \url{https://cds.cern.ch/record/2751915}.

\bibitem[Perepelitsa(2021)]{Perepelitsa:2751916}
Dennis Perepelitsa.
\newblock {sPHENIX contribution}.
\newblock Feb 2021.
\newblock URL \url{https://cds.cern.ch/record/2751916}.

\bibitem[Li(2021)]{Li:2752019}
Wei Li.
\newblock {STAR contribution}.
\newblock Feb 2021.
\newblock URL \url{https://cds.cern.ch/record/2752019}.

\bibitem[Bierlich(2021)]{Bierlich:2752020}
Christian Bierlich.
\newblock {A Pythia/Angantyr perspective on OO and pO collisions}.
\newblock Feb 2021.
\newblock URL \url{https://cds.cern.ch/record/2752020}.

\end{thebibliography}

\end{document}